# Participation is not a Design Fix for Machine Learning

Mona Sloane [\*1]   Emanuel Moss [\*2]   Olaitan Awomolo [\*3]   Laura Forlano [\*4]

## Abstract

This paper critically examines existing modes of participation in design practice and machine learning. Cautioning against "participation-washing", it suggests that the ML community must become attuned to possibly exploitative and extractive forms of community involvement and shift away from the prerogatives of context-independent scalability.

## 1. Introduction

Over the past years, we have seen mounting evidence of the disparate impact of ML systems on already oppressed and disadvantaged groups (Bolukbasi et al.; Buolamwini & Gebru; Eubanks; Noble; O'Neal). The experiences of oppression and privilege are structural challenges that are incredibly complex, and they are not new particularly not to the communities that suffer from them. But they have heightened alongside the exponential growth of wealth inequality alongside planetary destruction (Piketty; Hickel). It is therefore both unsurprising and promising that the ML community wishes to build "more democratic, cooperative, and participatory ML systems" (see workshop call).

Whilst this is an honorable goal, we want to caution against a familiar-sounding impulse towards "participation-washing" that we have seen in other areas of design and technology. For example, in the international development sector where "participation" of local communities at the receiving end of powerful agencies is based on manufactured consent and is based on (post-)colonial structures of global power (Peet & Hartwick); in the corporate sector where "users" are invited into "co-creation" sessions in order to create new product ideas; in the philanthropic sector where "the public" is challenged to join in defining new problems and/or solutions to "wicked problems"; or in the urban design or architecture sector where stakeholder engagement protocols often legitimize injustices in the (material) planning of space and systematically devalue user needs as part of profit- and scale-oriented design practices, or design inequality (Sloane, b;c).

## 2. Participatory Design

Participatory design methods can be traced to the 1970s when workers in Scandinavia worked together collaboratively to design the technologies that they would use in corporate settings (Schuler & Namioka; Sanders; Spinuzzi). Over the past several decades, participatory design and related concepts such as codesign and co-creation have been introduced as a way of engaging with ethics, values in design (Nissenbaum), value-sensitive design (Friedman 1996), and values levers in design (Shilton). Participatory design with its rich history in socially democratic countries in Europe, has sought to engage multiple stakeholders in deliberative processes in order to achieve consensus. At the same time, other approaches have emphasized agonism and the importance of dissensus, friction and disagreement (Keshavarz & Maze; DiSalvo, a; Mouffe; Hansson et al.). In this tradition of participatory design, the focus has been on designing publics (DiSalvo, b) to engage in matters of concern around complex socio-technical systems. In order to facilitate the engagement of multiple stakeholders in participatory design processes, designers often use prototypes, games (Flanagan & Nissenbaum) and other structured activities.

More recently, scholars have argued that nonhuman actors such as algorithms and machines (Choi et al.) as well as the multispecies (microbes, plants, animals and the natural environment) be considered as stakeholders in participatory design processes (Forlano & Halpern; Forlano, b; Heitlinger et al.). Finally, with the introduction of critical and speculative design and experiential futures in the early 2000s, design researchers have become interested in

[\*]Equal contribution  [1]Institute for Public Knowledge, New York University, New York, USA; Tübingen AI Center, BMBF Competence Centre for Machine Learning (TUE.AI). Eberhard Karls University of Tübingen, Tübingen, Germany. Mona Sloane's work was supported by the German Federal Ministry of Education and Research (BMBF): Tübingen AI Center, FKZ: 01IS18039A. [2]Data Society Research Institute, New York, USA; Department of Anthropology, CUNY Graduate Center, New York, USA [3]Temple University, Philadelphia, Pennsylvania [4]Illinois Institute of Technology, Chicago, Illinois. Correspondence to: Mona Sloane <ms11521@nyu.edu>.

*Proceedings of the 37$^{th}$ International Conference on Machine Learning*, Vienna, Austria, PMLR 119, 2020. Copyright 2020 by the author(s).

the ways in which participatory design and design futures might come together to create new modes of experiential futures (Candy; Candy & Dunagan), design fiction (Bleecker; Forlano & Mathew), speculative design (Dunne & Raby), speculative civics (DiSalvo et al.) and critical futures (Forlano & Halpern; Forlano, c) in order to think through the social consequences of emerging technologies.

Participatory design methods have often been seen as a way of overcoming supposed difficulties that users have in understanding ostensibly complex technologies, particularly in healthcare settings (Neuhauser & Kreps). Participatory methods have also been employed where designers anticipate public resistance or skepticism to a product or service (Asaro). The use of participatory methods in technology settings follows the development of participatory methods in other domains, particularly international development (Peet & Hartwick) where participation was seen as a means for overcoming local resistance to international development schemes (Goldman).

ML already incorporates certain forms of participation throughout the design of models and their integration into society, however participatory design practices from other domains hold important lessons for ML. We will expand the notion of "participation" beyond the forms of involvement that are commonly understood as participatory design. Following the review of key literature on participatory design and ML, we will introduce three different forms of participation: participation as work, participation as consultation, participation as justice, each illustrated with a list of examples. Through this framing, it becomes possible to understand how participatory design, a necessarily situated and context-dependent endeavor, articulates with industrial prerogatives of context-independent scalability. It also becomes possible to recognize where the discourse of participation fails to account for existing power dynamics and obscures the extractive nature of collaboration, openness, and sharing, particularly in corporate contexts. We conclude the paper with a set of recommendations drawn from considering a more expansive definition of participation in the context of ML.

## 3. Different Forms of Participation

### 3.1. Participation as Work

Much of ML plays out upon what is an intensely participatory field. Whether acknowledged or not, a broad range of participants play an important role in producing the data that is used to train and evaluate ML models. For example, ImageNet, which laid the foundations for deep learning and most image recognition applications and is still used for ML benchmarking, is an dataset of millions of images, taken by hundreds of thousands of people, scraped from the open web and labeled by mTurk workers (Krizhevsky et al.). Image classification tools are often built on top of models trained on the ImageNet dataset. Photographers, web designers, and mTurk workers all participate in every such application. A similar case presents itself for Natural Language Processing applications which, for over a decade, have sourced from Wikipedia for training language corpora (Gabrilovich & Markovitch).

Billions of ordinary web users also continually participate in the production and refinement of ML, as their online (and offline) activities produce neatly labeled rows of data on how they click their way around the web, navigate their streets, and engage in any number of other commercial, leisure, or romantic activities (Mayer-Schönberger & Cukier). Users also improve the performance of ML models as they interact with them, a single unanticipated click can update a model's parameters and future accuracy. This work sometimes is so deeply integrated into the ways in which users navigate the Internet that it is performed unconsciously, e.g. when using Google Maps and producing data movement patterns that enable traffic predictions. But other times it becomes more conscious, e.g. when classifying photos when completing a reCAPTCHA (O'Malley), or ranking Uber drivers (Rosenblat). Where ML technology does not live up to it's mythos, people work behind the veil to complete tasks as if by the magic of AI. Behind some mobile apps claiming to use AI are real people transcribing images of paper receipts and populating a purchase history database (Gray & Suri), moderating content (Roberts). The labor of integrating new technologies, such as AI applications, into everyday life and existing work processes and even out their rough edges, e.g. in healthcare (Sendak et al.), is the "human infrastructure" without which the socio-technical system cannot function (Elish & Mateescu). Labor, here, is multi-layered and includes affective and emotional labor, e.g. coping with stress and sleep-deprivation when integrating medical devices into everyday life (Forlano, a), or social labor, e.g. when explaining ML outcomes to users or even out their glitches such as when chatbots fail. All this work often happens without consent or acknowledgement, and remains uncompensated. Such ML design processes are cases of "designing for", i.e. processes that are void of a genuine integration of design users, relying on them to make the design product work ex post.

### 3.2. Participation as Consultation

In the case of participation as consultation, cf. (Martin Jr. et al., 2020), designers and technologists engage in episodic, short-term projects in which diverse stakeholders might be consulted at various stages of the process. This model is most common in architecture and urban planning as well as among major philanthropic foundations and private corporations. Architecture and urban planning practices use citizen participation approaches to engage different stakeholder groups in project development. As these projects are complex and have significant socio-economic impacts on communities, participatory workshops can provide an integrated framework where experts work with stakeholder groups to identify context-specific needs (Bratteteig & Verne; Saad-Sulonen & Horelli). Here, participation might be facilitated through small, face-to-face workshops or larger design sprints or hackathons as well as through the use of online platforms for crowdsourcing ideas.

There are several challenges that can limit the effectiveness of participation as consultation. For a variety of reasons including intellectual property concerns, in this model, long-term partnerships are either impossible, undesirable, unnecessary or cost prohibitive. As this type of top-down design process also takes the form of "designing for" a particular group without an ongoing commitment to their inclusion in the process, systemic inequalities that can be hard-coded into consultation and representation protocols (Sloane, c). Experts do not often have a good understanding of how to design effective participatory processes or engage the right stakeholders to achieve the desired outcomes. A third challenge occurs as cities begin to require participation workshops as part of the permitting and approvals process. Participation workshops can become performative, where experts do not actually take the needs or recommendations of the different stakeholder groups into consideration (Crosby et al.).

### 3.3. Participation as Justice

In the case of participation as justice, designers and technologists engage in more-long term partnerships with diverse stakeholders. In order to build trust, it is important to create ongoing relationships based on mutual benefit, reciprocity, equity and justice. Here, all members of the design process engage in more tightly coupled relationships with more frequent communication (which often happens through a blended communication and interaction approach, e.g. online/offline). The canon of participation as collaboration notably comprises participatory action research, which is focused on researchers and participants undertake action-oriented and self-reflexive practices that leads to them having more control over their lives (Baum et al.); infrastructuring, which centers designers' locations, the materials and systems intrinsic to designing, as well as (community) capacity building (Agid; Hillgren et al.; Le Dantec & DiSalvo); design justice, which goes beyond value-focused design and centers typically marginalized groups in collaborative and creative design processes that challenge and dismantle the matrix of domination, i.e. white supremacy, heteropatriarchy, capitalism, and settler colonialism (Costanza-Chock); crip technoscience, which refuses demands to eliminate disability, underscores that disabled people are expert designers of everyday life, and centers technoscientific activism, critical design practices, and disability justice (Hamraie & Fritsch); data feminism, which focuses on ideas of intersectional feminism (D'Ignazio & Klein); and tech activism and resistance, both from people affected by potentially harmful technology, such as the Atlantic Towers Resident Association in Brooklyn, NY (Gagne) and those designing it, see for example the Tech Worker Movement (Tarnoff), or a mix of both, such as Data for Black Lives, Black in AI, or LatinX in AI. What ties these approaches together is favoring using language around "designing with" in order to ensure that outcomes are valuable to people from diverse backgrounds and communities, including the disability community. Participation as justice has social and political importance, but it may be difficult to do it well, especially in a corporate context. Here, design justice can almost be seen as an oxymoron: given the extractive and oppressive capitalist logics and contexts of ML systems, it appears impossible to design ML products that are genuinely "just" and "equitable".

## 4. Critiques of Participation

The dominant mode of extraction within the ML industry is deeply entangled with the capitalist paradigm of scale, referring to the ability to gain revenue at a greater proportion per unit cost of inputs (Chandler & Hikino). But as a tech industry buzzword, the verb "to scale" refers to the ability of products to spread far beyond the context of development to new applications in new markets. Part of the promise of ML is that statistical generalizations learned from finite datasets will allow for inferences to be made across broader contexts, and that capabilities engendered by ML can be applied to additional settings without adding proportional costs. However, datasets are deeply context-bound, and that context, as well as the appropriateness of the use of those datasets, is lost in the scaling of ML applications (boyd & Crawford).

Acknowledging the modes of participation that are already components of ML challenges understandings of how these tools are able to scale. As such technologies scale across contexts, the generalizations that are learned inevitably require updating, by providing additional training data or correcting errors (Selbst et al.). This often requires the participation of users interacting with the system who experience the friction of providing additional information to the system (as with CAPTCHAs) or bearing the burden of system errors. As discussed above, representation/consultation is often prohibitively costly. Where a cost-benefit analysis may encourage such forms of participation in the earlier stages of product development, in later stages that product is expected to scale without incurring additional costs. The initial utility of representative and consultative forms of participation are thus diluted as products scale beyond the context in which that mode of participation contributed to the overall design of the product in earlier stages. For ML products to simultaneously scale and engage in meaningful partnerships oriented toward justice, they also require additional inputs of participation, and budgets must be set aside for that.

This can be thought of as levelling the playing field of futuring: product futures are often made very concrete for venture capitalists. But what kind of imaginative work do entrepreneurs do when it comes to the communities that they seek out as users (or targets) of their products? There is an existing imbalance between market-fit and community-fit. To address that and pave the way for design justice processes to become integral to ML, it is key to expand the notion of value beyond monetary value and the extractive logics underpinning the invasive data collection that is necessitated by most ML system designs. Promising developments have recently been made in the context of Indigenous data sovereignty which includes access, control and governance of Indigenous data (Anderson & Hudson).

Against that backdrop we suggest three cues for considering participation in ML in a more equitable way:

**1. Recognize participation as work.** Users already labor in, for, and through ML systems across a number of dimensions (affective, social, emotional). This labor upholds and improves ML systems and therefore is valuable for the owners of the ML systems. To acknowledge that, users should be asked for consent, be provided with opt-out options or alternatives, and, if they chose to participate through labor, be offered compensation. This could mean to clarify when and how data generated by user behaviour is used for the training and improvement of ML systems (e.g. via a banner on the Wikipedia page, or in Google Maps); to give an alternative security option for reCAPTCHA; to not punish users for refusing to leave reviews; to provide appropriate support for content moderators; to compensate "ghost workers" fairly (Gray & Suri); to develop reward systems for users that labor to integrate technologies into their lives and thereby provide rich data for profit-oriented ML companies.

**2. Participation as consultation must be designed for specific contexts.** If short-term participation is the most feasible and desired version for ML participation, then there needs to be a commitment to context-specificity, especially in terms of how the participation is facilitated. Every context is different, so participation has to be designed to address these different contexts. Rather than a one-size-fits-all approach, consultation and representation processes must be revisited and reexamined to ensure they are gathering the right information from the right people. As ML systems affect a wide range of groups, marginalized stakeholders should be given the space and voice to co-design and co-produce these systems (Crosby et al.). Documenting these processes and their contexts can form a knowledge base for long term, effective participation.

**3. Participation as justice must be genuine and long term.** This means to engage in creating processes that provide transparency and genuine knowledge sharing. This can be difficult particularly for proprietary design cases. Further, using the language of design justice without actually engaging in actual design justice processes and practices can only lead to corporate co-optation. For example, the ML field has seen a hype of "ethical AI" serving as a smokescreen for continuing with non-participatory and non-justice oriented ML design approaches (Sloane, a), despite good intentions. To avoid that, it may be helpful to make the tensions that characterize the goal of long term participation in ML visible, acknowledging that partnerships and justice do not scale in frictionless ways, but require constant maintenance and articulation with existing social formation in new contexts (Tsing).

We argue that it is crucial to enhance the ability for lateral thinking across applications and academic disciplines ("holistic futuring"), because harms can be produced by the same ways of thinking that produce the technology that causes the harms. This maps onto Vaughn's (Vaughan 1996) normalization of deviance and could benefit from cross-checking or lateral thinking between disciplines and forms of expertise. Such an approach could facilitate the development of an ontology of (design) harms or "design inequalities" (Sloane, a). To facilitate these efforts, we propose to develop a searchable database of design precedents across applications and disciplines that highlights design failures, especially failures of design participation, cross-referenced with socio-structural dimension (e.g. issues pertaining to racial inequality, or class-based inequity). This database should cover design projects across all sectors and domains, not just ML, and explicitly acknowledge deliberate absences and outliers which often are the most interesting and relevant social phenomena we can learn from (e.g. transgender identities). It may also acknowledge and educate on the deliberate refusal to "get counted" (D'Ignazio & Klein).

## 5. Conclusion

In this paper, we have cautioned against "participation-washing" of ML by critically examining the existing kinds of participation in design practice and ML. Existing forms of participation can be classified as work, as consultation, and as justice, but we have argued that the notion of "par-ticipation" should be expanded to acknowledge more subtle, and possibly exploitative, forms of community involvement in participatory ML design. This framing allows for understanding participatory design as a necessarily situated and context-dependent endeavor which is at odds with industrial prerogatives of extraction and context-independent scalability. Against that backdrop, it is imperative to recognize design participation as work; to ensure that participation as consultation is context-specific; and that participation as justice must be genuine and long term. Therefore, we argue that developing a cross-sectoral database of design participation failures that is cross-referenced with socio-structural dimensions and highlights "edge cases" that can and must be learned from.

**Participation is not a Design Fix for Machine Learning**